\documentclass[a4paper]{article}

\usepackage{arxiv}

\usepackage[utf8]{inputenc} 
\usepackage[T1]{fontenc}    
\usepackage{hyperref}       
\usepackage{url}            
\usepackage{booktabs}       
\usepackage{amsfonts}       
\usepackage{nicefrac}       
\usepackage{microtype}      
\usepackage{lipsum}		
\usepackage{graphicx}
\usepackage[numbers]{natbib}
\usepackage{doi}
\usepackage{latexsym}
\usepackage{amsmath}
\usepackage{amssymb}
\usepackage[mathscr]{eucal}
\usepackage{graphicx,color}
\usepackage{mathtools}
\usepackage{stmaryrd}
\usepackage{url}
\usepackage{xypic}

\usepackage{bm}

\usepackage{bbm}
\usepackage{tikz-cd}
\usepackage{psfrag}

\usepackage{subcaption} 

\newcommand{\R}{\mathbb{R}}
\newcommand{\SE}{\mathbb{SE}}
\newcommand{\SO}{\mathbb{SO}}

\newcommand{\SEthree}{\mathbb{SE}_3(3)}
\newcommand{\calI}{\mathcal{I}}
\newcommand{\calB}{\mathcal{B}}
\newcommand{\norm}[1]{\left\lVert #1\right\rVert}

\DeclareMathOperator{\diag}{diag}

\usepackage{amsthm}
\newtheorem{theorem}{Theorem}
\newtheorem{proposition}{Proposition}
\newtheorem{definition}{Definition}
\newtheorem{assumption}{Assumption}

\title{
A Geometric Observer for Visual--Inertial Odometry using Optical-Flow Constraints}

    \author{ \href{https://orcid.org/0000-0003-1116-7415}{\includegraphics[scale=0.06]{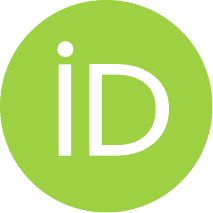}\hspace{1mm}Tarek Bouazza} \\
	I3S, CNRS, Université Côte d'Azur\\
        06900 Sophia Antipolis, France \\
	\texttt{bouazza@i3s.unice.fr} \\
	\And
     \href{https://orcid.org/0000-0002-8718-8004}{\includegraphics[scale=0.06]{orcid.pdf}\hspace{1mm}Soulaimane Berkane} \\
        Département d’informatique et d’ingénierie, \\
	Université du Québec en Outaouis \\
        and Department of Electrical Engineering, \\ 
        Lakehead University \\
        Gatineau, QC J8X 3X7, Canada \\
	\texttt{soulaimane.berkane@uqo.ca} \\
     \And
	\href{https://orcid.org/0000-0002-7779-1264}{\includegraphics[scale=0.06]{orcid.pdf}\hspace{1mm}Tarek Hamel} \\
    I3S, CNRS, Université Côte d'Azur\\
    and Institut Universitaire de France (IUF) \\
    06900 Sophia Antipolis, France \\
	\texttt{thamel@i3s.unice.fr} \\
}

\date{}

\renewcommand{\headeright}{}
\renewcommand{\undertitle}{A preprint\thanks{Under Review}}

\renewcommand{\shorttitle}{}

\hypersetup{
pdftitle={A template for the arxiv style},
pdfsubject={q-bio.NC, q-bio.QM},
pdfauthor={David S.~Hippocampus, Elias D.~Striatum},
pdfkeywords={First keyword, Second keyword, More},
}

\begin{document}
\maketitle

\begin{abstract}
This paper presents a geometric observer for monocular visual--inertial odometry driven by individual optical-flow constraints. For each tracked point, the bearing and its rotationally compensated derivative give a depth-free homogeneous constraint on the body-frame velocity. These constraints enter the innovation of an observer on the gravity-augmented group $\SEthree$ through a Riccati gain, without estimating depths or reconstructing the direction of motion. The auxiliary gravity state makes the six-dimensional body-frame velocity--gravity error dynamics independent of the attitude correction, and their uniform observability is shown to be equivalent to persistent excitation of a matrix formed from the inertial bearings and their derivatives. Under this condition, the body-frame velocity and gravity errors converge globally and uniformly exponentially, and a gravity-based attitude correction renders the combined velocity, gravity, and reduced-attitude error almost globally asymptotically stable and locally uniformly exponentially stable. Simulations illustrate the excitation condition, recovery from initial errors, and sensitivity to perturbed measurements.
\end{abstract}

\section{Introduction}

Visual--inertial odometry (VIO) estimates the motion of a platform from camera and inertial measurements, typically where global positioning is unavailable. Most current systems rely on extended Kalman filtering, as in MSCKF \cite{mourikis2007multi} and ROVIO \cite{bloesch2015robust}, or on nonlinear optimization over a sliding window, as in OKVIS \cite{leutenegger2013keyframe} and VINS-Mono \cite{qin2018vins}. Nonlinear observers complement these methods with convergence guarantees that do not rely on linearization about the true state.

Geometric observers exploit the symmetries of the navigation model to define estimation errors whose dynamics can be analyzed on the underlying state space \cite{barrau2016invariant,mahony2017geometric,van2023eqvio,fornasier2023msceqf,van2023synchronous}. A central difficulty in inertial navigation is the coupling between attitude and translational estimation: the attitude resolves the measured specific acceleration in the inertial frame, while translational measurements are used to correct the attitude \cite{berkane2021navigation}. Auxiliary states can remove the attitude error from the translational error dynamics, as shown for vision-aided inertial navigation with known landmarks in \cite{wang2022nonlinear} and for inertial and body-frame measurements in \cite{benahmed2025geometric}. For unknown landmarks, Boughellaba et al.\ embed the attitude, velocity, position, an auxiliary gravity vector, and the landmarks in the matrix Lie group $\SE_{3+n}(3)$, first for landmark--inertial SLAM \cite{boughellaba2025landmark} and then for VIO, including monocular bearing measurements \cite{boughellaba2026nonlinear}. The VIO formulation couples a Riccati translational observer to a geometric attitude observer and retains the landmarks as states.

Optical-flow offers a way to avoid landmark states. For a static point, the bearing and its time derivative constrain the instantaneous camera velocity without involving the depth of the point. In our previous work \cite{bouazza2025observer}, these constraints were combined by an iterative optimization on $\mathbb S^2$ to reconstruct the direction of the body-frame velocity, which was then supplied as a measurement to a Riccati observer for the body-frame velocity and gravity. The observer therefore depends on an intermediate optimization and receives only the reconstructed direction, not the individual constraints.

We propose a formulation of optical-flow-based VIO on the gravity-augmented navigation group $\SEthree$ and an observer driven directly by the individual bearing-flow constraints. The group state combines attitude, gravity, velocity, and position, without introducing a state for each observed landmark. The auxiliary gravity column separates the body-frame velocity--gravity error dynamics from the attitude correction, while the visual constraints provide the implicit outputs used to correct this subsystem. The contributions are threefold: (i) a group observer driven by the individual bearing-flow constraints, with state and Riccati dimensions independent of the number of features; (ii) a necessary and sufficient excitation condition for uniform observability of the velocity--gravity error system, interpreted in terms of the motion and the scene; (iii) global uniform exponential stability of the velocity--gravity error and almost-global asymptotic stability of the combined error with the reduced attitude.

\medskip\noindent\textbf{Notation.} The sets of real and nonnegative real numbers are denoted by $\R$ and $\R_+$, and $\R^n$ is the $n$-dimensional Euclidean space.
The operator $\operatorname{col}(\cdot)$ stacks its arguments vertically.
The notation $\norm{\cdot}$ denotes the Euclidean norm for vectors and its induced norm for matrices.
We denote by $\R^ {m \times n}$ the set of real $m \times n$ matrices and by $I_n$ the $n\times n$ identity matrix. For a symmetric matrix $P$, $P\succ0$ and $P\succeq0$ denote positive definiteness and positive semidefiniteness, respectively, and $\lambda_{\min}(P)$ denotes its smallest eigenvalue. The Kronecker product is denoted by $\otimes$.
The unit sphere is $\mathbb S^2:=\{u\in\R^3:\norm{u}=1\}$, and $\SO(3):=\{R\in\R^{3\times3}:R^\top R=I_3,\ \det R=1\}$ is the rotation group.
For $u\in\R^3$, the skew-symmetric matrix $u^\times$ satisfies $u^\times w=u\times w$, and for $u\neq0$, $\pi_u:=I_3-uu^\top/\norm{u}^2$ is the projection onto the plane orthogonal to $u$.

\section{VIO problem formulation}\label{sec:problem}

\begin{figure}[t]
    \centering
    \includegraphics[scale=1.25]{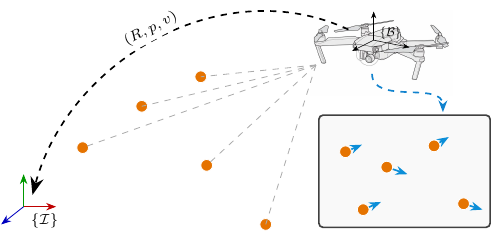}
    \caption{Optical-flow-based VIO. The onboard camera observes static landmarks, whose apparent image motion is estimated by optical-flow, while the IMU provides the vehicle's inertial measurements.} 
    \label{fig:vio-framework}
\end{figure}

We consider a rigid body carrying an inertial measurement unit (IMU) and a monocular camera that observes a static scene, as illustrated in Fig.~\ref{fig:vio-framework}. Let $\{\calI\}$ and $\{\calB\}$ denote the inertial and body frames, and let $R\in\SO(3)$ map body-frame coordinates to inertial-frame coordinates. The inertial position and velocity are $p,v\in\R^3$, and the constant gravity vector is $g=g_0e_3$, with $g_0\approx9.81\,\mathrm{m/s^2}$ and vertical direction $e_3\in\mathbb S^2$. The IMU provides the angular velocity $\omega\in\R^3$ and the specific acceleration $a\in\R^3$, both expressed in the body frame. The calibrated camera frame coincides with the body frame, including its origin. The kinematics are
\begin{align}
    \dot R&=R\omega^\times, &
    \dot p&=v, &
    \dot v&=g+Ra, &
    \dot g&=0. \label{eq:dynamics}
\end{align}

\subsection{Lie-group representation of the extended navigation state}
The choice of state representation is motivated by the coupling between attitude and translational estimation. The visual constraints involve the body-frame velocity $v_\calB:=R^\top v$, whose propagation requires the body-frame gravity $g_\calB:=R^\top g$. If gravity is treated only as a known inertial vector, the natural velocity propagation uses $\hat R^\top g$, where $\hat R$ denotes the attitude estimate (hats denote estimates throughout), and its error contains $(R-\hat R)^\top g$. Correcting the attitude from the same visual information then gives a feedback interconnection between the two estimation errors~\cite{benahmed2025geometric,boughellaba2026nonlinear}.

To obtain a decoupled error system, we include gravity in the estimated navigation state. This follows the auxiliary-state approach of~\cite{wang2022nonlinear,benahmed2025geometric} and the gravity augmentation used in~\cite{boughellaba2025landmark,boughellaba2026nonlinear}. Although the true gravity vector remains equal to the known constant $g$, its estimate $\hat g\in\R^3$ is carried as a separate observer state. The observer can therefore correct velocity and gravity jointly from the visual constraints, and use the resulting gravity estimate to correct attitude. The error analysis in Section~\ref{sec:analysis} will show precisely how this additional state removes the attitude error from the velocity--gravity dynamics. Specifically, we use the matrix group
\begin{equation}\label{eq:SE3definition}
\SEthree:=\left\{\begin{bmatrix}R&Z\\0&I_3\end{bmatrix}:R\in\SO(3),\ Z\in\R^{3\times3}\right\}.
\end{equation}
Writing a group element as $(R,Z)$, multiplication and inversion are $(R,Z)(R',Z')=(RR',Z+RZ')$ and $(R,Z)^{-1}=(R^\top,-R^\top Z)$. Its Lie algebra consists of the matrices
\begin{equation}\label{eq:algebramap}
 \mathcal L(\xi,z_g,z_v,z_p):=
 \begin{bmatrix}
 \xi^\times&\begin{matrix}z_g&z_v&z_p\end{matrix}\\
 0_{3\times3}&0_{3\times3}
 \end{bmatrix},
\end{equation}
with $\xi,z_g,z_v,z_p\in\R^3$. We collect the navigation variables into the extended pose matrix
\begin{equation}\label{eq:groupstate}
    X:=\mathcal T_3(R,p,v,g):=
    \begin{bmatrix}
        R&\begin{matrix}g&v&p\end{matrix}\\
        0_{3\times3}&I_3
    \end{bmatrix}\in\SEthree.
\end{equation}
The column ordering in \eqref{eq:groupstate} follows the integrator chain $\dot g=0$, $\dot v=g+Ra$, and $\dot p=v$. Accordingly, the dynamics \eqref{eq:dynamics} take the group-affine form \cite{barrau2016invariant}
\begin{equation}\label{eq:groupdynamics}
    \dot X=[X,D]+XU.
\end{equation}
Here $[X,D]:=XD-DX$ denotes the matrix commutator and $D:=\diag(0_{3\times3},D_0)$, where
\begin{equation}\label{eq:groupinputs}
 U:=\mathcal L(\omega,0,a,0),\qquad
 D_0:=\begin{bmatrix}0&1&0\\0&0&1\\0&0&0\end{bmatrix}.
\end{equation}
The product $XU$ accounts for the measured angular velocity and specific acceleration, while the commutator $[X,D]$ supplies the remaining terms of the integrator chain by introducing $g$ into the velocity column and $v$ into the position column.

\subsection{Per-feature optical-flow constraints}
To define the measurements, let $P_i\in\R^3$, $i=1,\ldots,n$, denote unknown stationary scene points. The relative position and unit bearing of $P_i$ in the body frame are
\begin{equation}\label{eq:bearing}
    Y_i:=R^\top(P_i-p),\qquad b_i:=\frac{Y_i}{\norm{Y_i}}\in\mathbb S^2.
\end{equation}
Since $\dot Y_i=-\omega^\times Y_i-v_{\cal B}$ and $\dot b_i=\pi_{b_i}\dot Y_i/\norm{Y_i}$, we obtain
\begin{equation}\label{eq:flow}
    \dot b_i=-\frac{1}{\norm{Y_i}}\pi_{b_i}v_{\cal B}-\omega^\times b_i.
\end{equation}
Equation~\eqref{eq:flow} describes the spherical motion field induced by the camera motion, which characterizes the velocity of the bearing of a static scene point.
Optical-flow methods estimate the corresponding image-plane motion from successive frames~\cite{horn1981determining}, and the calibrated projection onto the bearing sphere maps this estimate to a tangent vector at $b_i$. An estimate of the bearing derivative is thus obtained from the image sequence itself, without separately differentiating a bearing history. 
This estimate approximates the motion field as long as the tracked image region stays on the same scene point, and may degrade under appearance changes or occlusions. The analysis below assumes the exact motion field \eqref{eq:flow}.

We use the gyroscope measurement to remove the rotational part of this flow. The compensated flow $s_i:=\dot b_i+\omega^\times b_i=-\pi_{b_i}v_{\cal B}/\norm{Y_i}$ depends on the unknown range only through the positive factor $1/\norm{Y_i}$, which the cross product with $s_i$ eliminates:
\begin{equation}\label{eq:individualconstraint}
    s_i^\times\pi_{b_i}v_{\cal B}=0,\qquad i=1,\ldots,n.
\end{equation}
To express these constraints in terms of the group state, we introduce the row selector $J:=[I_3\;0_{3\times3}]$ and the column selectors $e_g,e_v,e_p\in\R^6$, defined as the fourth, fifth, and sixth standard basis vectors, respectively. Thus $JXe_g=g$, $JXe_v=v$, and $JXe_p=p$, while the inverse state contains the corresponding body-frame quantities with a minus sign. Since $JX^{-1}e_v=-v_\calB$, \eqref{eq:individualconstraint} becomes the implicit output
\begin{equation}\label{eq:groupoutput}
 h_i(t,X):=H_i(t)JX^{-1}e_v=0,\qquad
 H_i:=s_i^\times\pi_{b_i}.
\end{equation}
The coefficient $H_i(t)$ is computed from measured bearing and flow. Thus \eqref{eq:groupoutput} constrains the inverse column $X^{-1}e_v$, although neither this column nor the depth is measured.

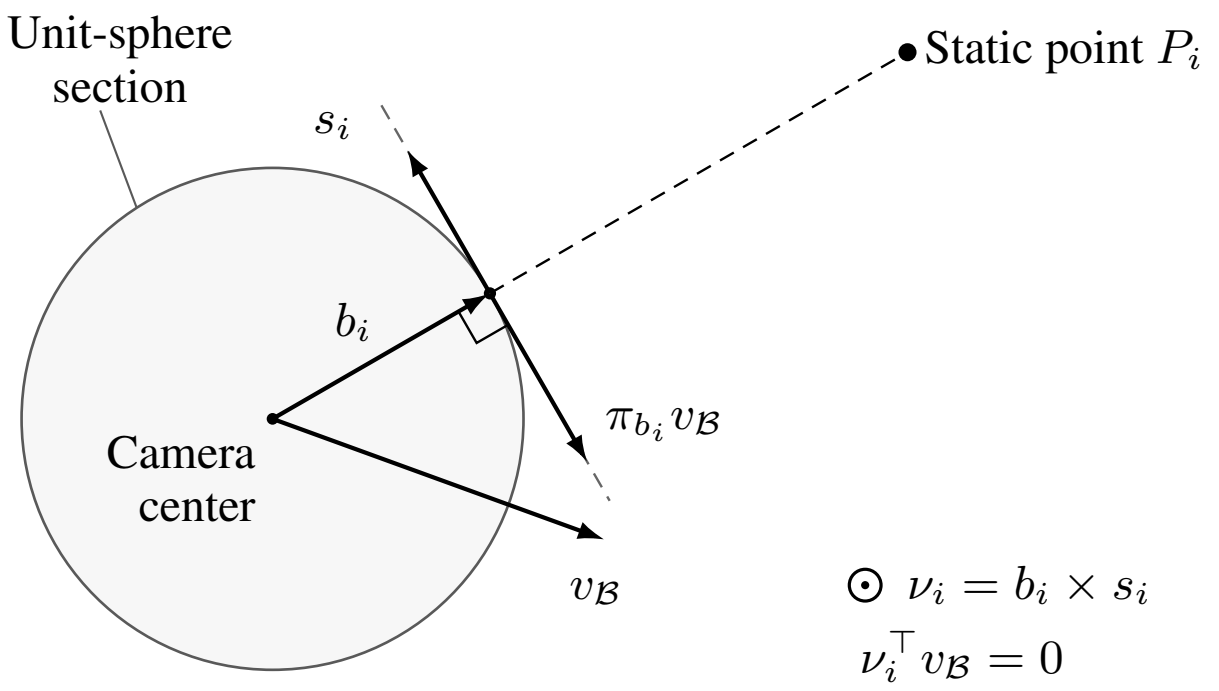
\begin{figure}[t]
    \centering
\begin{tikzpicture}[
    x=1cm,y=1cm,
    font=\footnotesize,
    draw=black,text=black,
    >={Latex[length=1.8mm,width=1.2mm]},
    vector/.style={->,line width=0.8pt},
    guide/.style={densely dashed,line width=0.45pt},
    labelbox/.style={fill=white,inner sep=1.5pt,align=center}
]
\path[use as bounding box] (0,0.15) rectangle (8.35,4.85);
\coordinate (O) at (2.00,2.00);
\coordinate (B) at (3.438,2.830);
\coordinate (F) at (6.20,4.425);
\coordinate (V) at (4.20,1.20);
\coordinate (S) at (2.888,3.783);
\coordinate (T) at (4.084,1.711);

\fill[black!3] (O) circle[radius=1.66];
\draw[line width=0.5pt,black!65] (O) circle[radius=1.66];
\node[align=center,anchor=west] at (0.12,4.38)
    {Unit-sphere\\section};
\draw[line width=0.4pt,black!65] (0.86,4.04) -- (1.10,3.40);

\draw[guide] (B) -- (F);
\draw[vector] (O) -- (B)
    node[pos=0.52,above left,inner sep=2pt] {$b_i$};
\fill (O) circle[radius=1.1pt];
\node[anchor=north east,align=right,inner sep=3pt] at (O)
    {Camera\\center};
\fill (B) circle[radius=1.15pt];
\fill (F) circle[radius=1.7pt];
\node[anchor=west,align=left,inner sep=3pt] at (F)
    {Static point $P_i$};

\draw[guide,black!60] (2.72,4.074) -- (4.23,1.458);
\draw[vector] (O) -- (V);
\node at (4.14,0.86) {$v_{\cal B}$};
\draw[vector] (B) -- (S);
\node at (2.40,3.94) {$s_i$};
\draw[vector] (B) -- (T);
\node[labelbox,anchor=west] at (4.14,1.96) {$\pi_{b_i}v_{\cal B}$};

\draw[line width=0.45pt]
    (3.230,2.710) -- (3.350,2.502) -- (3.558,2.622);

\draw[line width=0.6pt] (5.92,0.90) circle[radius=0.105];
\fill (5.92,0.90) circle[radius=0.027];
\node[anchor=west,inner sep=4pt] at (6.05,0.90) {$\nu_i=b_i\times s_i$};
\node[anchor=west] at (5.76,0.43) {$\nu_i^\top v_{\cal B}=0$};
\end{tikzpicture}
    \caption{Depth-free constraint from one feature with $s_i\neq0$, viewed in the plane $\nu_i^\perp$. The circle is a section of the unit bearing sphere; the dashed tangent contains both $s_i$ and $\pi_{b_i}v_{\cal B}$, which point in opposite directions.}
    \label{fig:flow-geometry}
\end{figure}

Since $b_i^\top s_i=0$, the coefficient $H_i$ satisfies
\begin{equation}\label{eq:rankoneflow}
 H_i=b_i\nu_i^\top,\qquad \nu_i:=b_i\times s_i,\qquad
 H_i^\top H_i=\nu_i\nu_i^\top.
\end{equation}
Hence $H_i$ has rank one when $s_i\neq0$, and \eqref{eq:groupoutput} reduces to the scalar constraint $\nu_i^\top v_{\cal B}=0$. Geometrically, $v_{\cal B}=(b_i^\top v_{\cal B})b_i-\norm{Y_i}s_i$ lies in $\operatorname{span}\{b_i,s_i\}$, whose normal is $\nu_i$ (Fig.~\ref{fig:flow-geometry}). A feature with $s_i=0$ contributes a zero block. Because the constraints are homogeneous in $v_\calB$, the speed is undetermined at every instant; it can be recovered only through the inertial dynamics and the variation of the constraints over time.

\begin{assumption}\label{ass:standing}
The true trajectory of \eqref{eq:dynamics} is defined for all $t\in\R_+$, with $\omega$ and $a$ continuous and $\omega$ bounded. The tracked set $\{P_1,\ldots,P_n\}$ is fixed, and each point $P_i$ is stationary and satisfies $\norm{P_i-p(t)}>0$ for all $t\in\R_+$. The compensated flows $s_i$ are bounded. The measurements are calibrated, synchronized, and noise free, and the IMU biases are zero.
\end{assumption}

\subsection{VIO invariance and group symmetries}

The visual and inertial measurements do not determine the origin of the inertial frame or its orientation about gravity~\cite{li2013high,boughellaba2026nonlinear}. To express this ambiguity in the group representation, consider a change of reference consisting of a translation $c\in\R^3$ and a rotation $Q\in\SO(3)$ satisfying $Qe_3=e_3$. The latter condition ensures that the known gravity vector is preserved, since $Qg=g$. We associate this change of reference with the group element
\begin{equation}\label{eq:groupgauge}
    \mathcal G_{Q,c}:=
    \begin{bmatrix}
        Q&\begin{matrix}0&0&c\end{matrix}\\
        0_{3\times3}&I_3
    \end{bmatrix}\in\SEthree.
\end{equation}
Its left action on the navigation state is $\mathcal G_{Q,c}X=\mathcal T_3(QR,Qp+c,Qv,g)$. Since $[\mathcal G_{Q,c},D]=0$, the transformed state satisfies \eqref{eq:groupdynamics} with the same input $U$. Writing $R':=QR$ and $p':=Qp+c$, and transforming the unknown landmarks as $P_i':=QP_i+c$, gives $R'^\top(P_i'-p')=R^\top(P_i-p)$, so the bearings and flows are unchanged. These transformations form the gauge group $\mathcal S:=\{\mathcal G_{Q,c}:Q\in\SO(3),\ Qe_3=e_3,\ c\in\R^3\}\subset\SEthree$,
since $\mathcal G_{Q,c}\mathcal G_{Q',c'}=\mathcal G_{QQ',c+Qc'}$. Every state in $\mathcal SX$ produces identical measurements when the unknown scene is transformed accordingly. Position and yaw are therefore not observable. The body-frame velocity $v_\calB$, the body-frame gravity $g_\calB$, and the reduced attitude $R^\top e_3$ are invariant under $\mathcal S$, and the convergence results below concern these quantities.
Their observability depends on the motion and the scene (Section~\ref{sec:analysis}).

\section{Observer design on \texorpdfstring{$\SEthree$}{SE3(3)}}\label{sec:observer}

\begin{figure}[t]
 \centering
\begingroup
\begin{tikzpicture}[
  x=1cm,y=1cm,font=\fontsize{9}{9.5}\selectfont,text=black,
  block/.style={draw=black,line width=0.55pt,rounded corners=0.7mm,
    fill=black!3,align=center,inner xsep=3pt,inner ysep=4pt,outer sep=0.4pt},
  flow/.style={draw=black,line width=0.6pt,-{Latex[length=1.6mm,width=1.1mm]}},
  bus/.style={draw=black,line width=0.6pt},
  signal/.style={font=\fontsize{8}{9.5}\selectfont,fill=white,inner sep=1pt}
]
  \path[use as bounding box] (0,-0.85) rectangle (8.45,7.05);
  \node[block,fill=white,minimum width=3.55cm,minimum height=0.72cm]
    (camera) at (2.05,6.85) {\textbf{Camera and optical flow}\\$b_i,\dot b_i$};
  \node[block,fill=white,minimum width=3.45cm,minimum height=0.72cm]
    (imu) at (6.25,6.85) {\textbf{IMU}\\$\omega,a$};
  \node[block,minimum width=3.55cm,minimum height=0.95cm]
    (flow) at (2.05,5.15)
    {\textbf{Flow coefficients}\\$s_i=\dot b_i+\omega^\times b_i$\\$H_i=s_i^\times\pi_{b_i}$};
  \node[block,minimum width=3.45cm,minimum height=0.95cm]
    (gain) at (6.25,5.15)
    {\textbf{Riccati gain}\\$L=PC^\top W_o$\\$A(\omega),\ C(H_1,\ldots,H_n)$};
  \draw[flow] (camera.south) -- (flow.north);
  \draw[flow] (imu.south) -- (gain.north);
  \draw[flow] (imu.south) -- (6.25,6.1) --
    node[signal,above,pos=0.55] {$\omega$} (3.15,6.1) -- (3.15,5.76);
  \draw[flow] (flow.east) -- node[signal,above] {$H_i$} (gain.west);
  \node[block,minimum width=3.55cm,minimum height=0.82cm]
    (residual) at (2.05,3.6)
    {\textbf{Group output residual}\\$y_i=H_iJ\hat X^{-1}e_v$};
  \draw[flow] (flow.south) -- (residual.north);
  \node[block,minimum width=7.20cm,minimum height=1.12cm]
    (innovation) at (4.15,1.8)
    {\textbf{Lie-algebra innovation}\\
     $\Lambda=\mathcal L(-\sigma_R,K_g y,K_v y,0)$\\[.2em]
     $\sigma_R=(k_g/g_0^2)\,g\times J\hat Xe_g$};
  \draw[flow] (residual.south) --
    node[signal,right] {$y$} (residual.south |- innovation.north);
  \draw[flow] (gain.south) --
    node[signal,right] {$L$}
    (gain.south |- innovation.north);
  \node[block,fill=black!7,line width=0.75pt,
    minimum width=7.20cm,minimum height=0.88cm]
    (state) at (4.15,-0.15)
    {\textbf{Group state estimate}\quad $\hat X\in\SEthree$\\[.2em]
     $\dot{\hat X}=[\hat X,D]+\hat XU+\Lambda\hat X$};
  \draw[flow] (innovation.south) -- node[signal,right] {$\Lambda$} (state.north);
  \draw[flow] (imu.east) -- (8.45,6.85) --
    node[signal,left,pos=0.72] {$U$} (8.45,-0.15) -- (state.east);
  \draw[bus] (state.west) -- (-0.08,-0.15) --
    node[signal,right,pos=0.26] {$\hat X$} (-0.08,3.6);
  \draw[flow] (-0.08,3.6) -- (residual.west);
  \draw[flow] (-0.08,1.8) -- (innovation.west);
  \fill[black] (0.08,1.75) circle (0.7pt);
\end{tikzpicture}
\endgroup
 \caption{Observer construction on $\SEthree$. The measured flow coefficients and the estimated inverse group column form the visual residual. The residual, the Riccati gain, and the estimated gravity column determine the Lie-algebra innovation, which corrects the group state.}
 \label{fig:observer-architecture}
\end{figure}
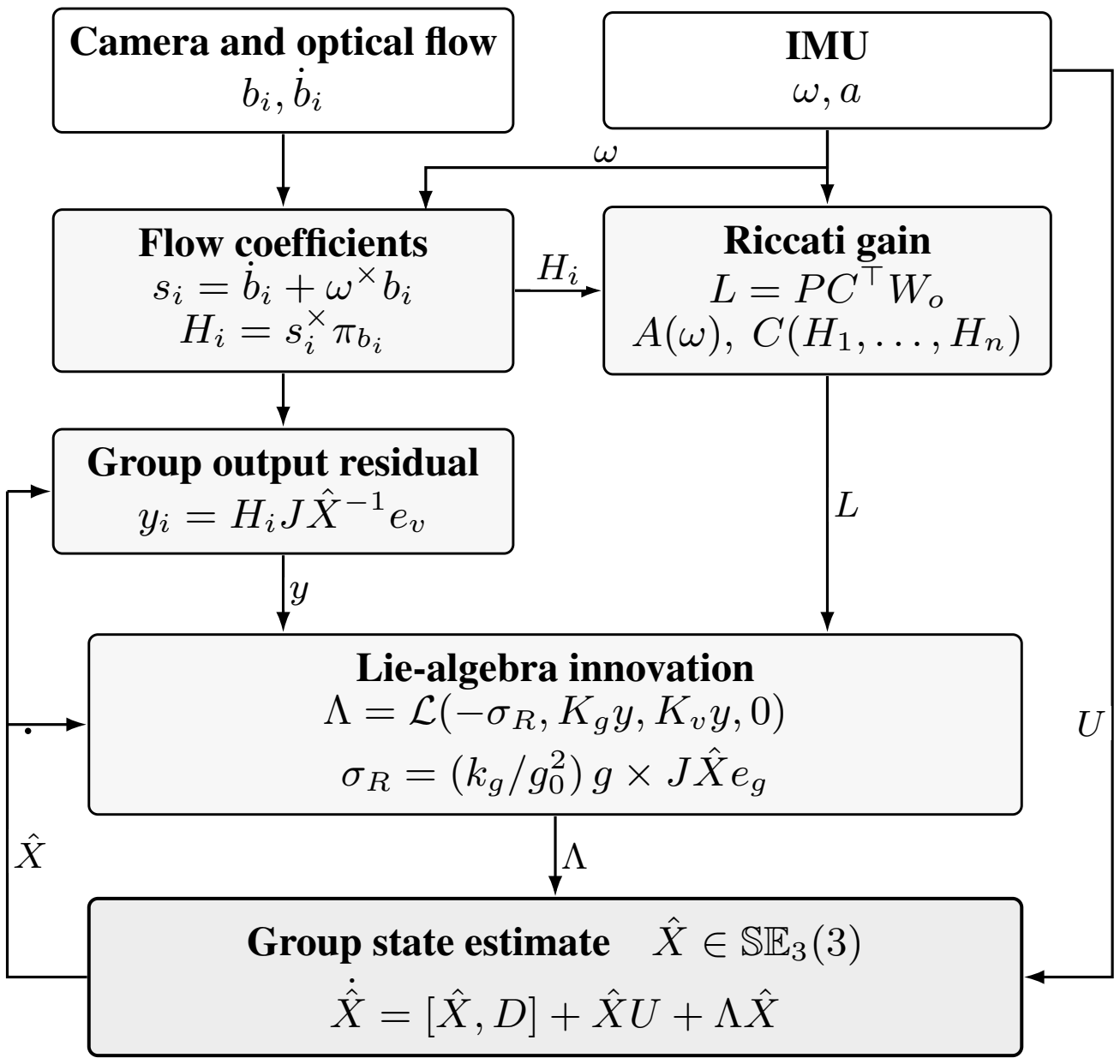

The observer copies the group dynamics \eqref{eq:groupdynamics} and adds an innovation built from the visual constraints and the known gravity vector. The visual part of the innovation is used to correct the velocity and auxiliary gravity estimates, while the attitude correction compares the resulting gravity estimate with the known gravity direction. Figure~\ref{fig:observer-architecture} illustrates the overall observer structure.
Let $\hat X=\mathcal T_3(\hat R,\hat p,\hat v,\hat g)\in\SEthree$ denote the estimated state. Since the true state satisfies the implicit outputs \eqref{eq:groupoutput}, evaluating these outputs at $\hat X$ gives the residuals $y_i:=h_i(t,\hat X)=H_iJ\hat X^{-1}e_v$, stacked as $y:=\operatorname{col}(y_1,\ldots,y_n)$; each measures the violation of one feature constraint and vanishes when $\hat X=X$. 
The group observer is
\begin{equation}\label{eq:groupobserver}
 \dot{\hat X}=[\hat X,D]+\hat XU+\Lambda\hat X,
\end{equation}
where the innovation is defined by the Lie-algebra element
\begin{equation}\label{eq:Delta}
 \Lambda:=\mathcal L(-\sigma_R,K_g y,K_v y,0).
\end{equation}
The gains $K_v,K_g\in\R^{3\times3n}$ are constructed below from a Riccati equation for the velocity--gravity subsystem. The position column of $\Lambda$ is zero, so the position estimate receives no visual correction of its own. 

The rotational part compares the estimated gravity column $J\hat Xe_g=\hat g$ with the known gravity through
\begin{equation}\label{eq:attitudeobserver}
 \sigma_R:=\frac{k_g}{g_0^2}\,g\times J\hat Xe_g,\qquad k_g>0.
\end{equation}
This is the single-direction complementary correction of~\cite{mahony2008nonlinear}, in which the body-frame vector $\hat R^\top\hat g$ produced by the translational part of the observer replaces a gravity measurement. The gravity estimate is redundant, since the observer does not explicitly impose either $\hat g=g$ or $\norm{\hat g}=g_0$. The convergence of the body-frame estimate $\hat R^\top\hat g$ to $R^\top g$ is established through the velocity--gravity error analysis in Section~\ref{sec:analysis}. 
The visual gains are constructed using the following measured-signal matrices:
\begin{equation}\label{eq:Coutput}
 A(t):=\begin{bmatrix}-\omega^\times&I_3\\0&-\omega^\times\end{bmatrix},\quad
 C(t):=\operatorname{col}(H_1,\ldots,H_n)\,C_*,
\end{equation}
with $C_*:=[I_3\;0_{3\times3}]$. The matrix $A(t)$ describes the velocity--gravity propagation in body coordinates, whereas $C(t)$ collects the visual constraints on the velocity component. As shown in Section~\ref{sec:analysis}, these matrices also characterize the exact estimation-error dynamics. Since both matrices depend only on measured signals, the Riccati equation
\begin{equation}\label{eq:CRE}
 \dot P=AP+PA^\top-PC^\top W_oCP+S,\qquad P(0)=P_0,
\end{equation}
can be integrated alongside the observer. Here $W_o\in\R^{3n\times3n}$ and $S\in\R^{6\times6}$ are constant symmetric positive-definite design weights, and $P_0\succ0$ does not depend on the initial estimate. The gains in \eqref{eq:Delta} are then
\begin{equation}\label{eq:gainlift}
 \operatorname{col}(K_v,K_g):=(I_2\otimes\hat R)\,L,\qquad L:=PC^\top W_o.
\end{equation}
The Riccati gain $L$ acts in body coordinates, its first three rows correcting the velocity and its last three rows the gravity. The factor $I_2\otimes\hat R$ maps these corrections into the inertial frame, in which the translational columns of $\Lambda$ are expressed. Consequently, $\operatorname{col}(\hat R^\top K_vy,\hat R^\top K_gy)=Ly$, and the visual correction in body coordinates is precisely the Riccati correction.

\section{Closed-loop analysis}\label{sec:analysis}
We first derive the estimation-error dynamics, obtaining a linear time-varying subsystem for the body-frame velocity and gravity errors. We then characterize its uniform observability in terms of the motion and the scene, and use its exponential convergence to study the reduced attitude.

\subsection{Velocity--gravity error dynamics and convergence}

To compare the true and estimated group states, define the right-invariant error $E:=X\hat X^{-1}$. Along \eqref{eq:groupdynamics} and \eqref{eq:groupobserver}, the common input $U$ cancels from its derivative, and the left innovation appears as a right factor:
\begin{equation}\label{eq:group-error-dynamics}
 \dot E=[E,D]-E\Lambda.
\end{equation}
Writing $E=\mathcal T_3(\tilde R,\tilde p,\tilde v,\tilde g)$, we have $\tilde R=R\hat R^\top$, $\tilde g=g-\tilde R\hat g$, $\tilde v=v-\tilde R\hat v$, and $\tilde p=p-\tilde R\hat p$, and \eqref{eq:group-error-dynamics} yields
\begin{equation}\label{eq:groupcolumns}
 \dot{\tilde R}=\tilde R\sigma_R^\times,\quad
 \dot{\tilde g}=-\tilde R K_g y,\quad
 \dot{\tilde v}=\tilde g-\tilde R K_v y,\quad
 \dot{\tilde p}=\tilde v.
\end{equation}
Since $\Lambda e_g=\operatorname{col}(K_gy,0_3)$ and $\Lambda e_v=\operatorname{col}(K_vy,0_3)$, $\sigma_R$ does not enter the equations for $\tilde g$ and $\tilde v$. 
The auxiliary gravity state is essential for this decoupling. Indeed, using the known gravity vector $g$ directly instead of an estimated gravity state would yield
$\dot{\tilde v}=(I_3-\tilde R)g-\tilde RK_vy$, which depends on the attitude error. To relate these equations to the body-frame visual residuals, define $\hat v_\calB:=\hat R^\top\hat v$ and $\hat g_\calB:=\hat R^\top\hat g$. The inverse group columns then give the error coordinates
\begin{equation}\label{eq:xdef}
 x:=\begin{bmatrix}JX^{-1}(E-I_6)e_v\\JX^{-1}(E-I_6)e_g\end{bmatrix}
   =\begin{bmatrix}R^\top\tilde v\\R^\top\tilde g\end{bmatrix}
   =:\begin{bmatrix}x_v\\x_g\end{bmatrix}.
\end{equation}
Equivalently, $x_v=v_{\cal B}-\hat v_{\cal B}$ and $x_g=g_{\cal B}-\hat g_{\cal B}$. Since $\hat X^{-1}=X^{-1}E$ and $h_i(t,X)=0$, the residual satisfies $y_i=H_iJX^{-1}(E-I_6)e_v=H_i x_v$, so that $y=C(t)x$, with $C$ defined in \eqref{eq:Coutput}. Differentiating \eqref{eq:xdef} and using \eqref{eq:groupcolumns} gives $\dot x=A(t)x-\operatorname{col}(\hat R^\top K_vy,\hat R^\top K_gy)$, with $A$ from \eqref{eq:Coutput}, and $\operatorname{col}(\hat R^\top K_vy,\hat R^\top K_gy)=Ly$ by \eqref{eq:gainlift}; hence
\begin{equation}\label{eq:closederror}
 \dot x=A(t)x-Ly=(A-PC^\top W_oC)x,
\end{equation}
which is exact and independent of the attitude error. In components, \eqref{eq:groupobserver} reads
\begin{equation}\label{eq:observercomponents}
\begin{aligned}
 \dot{\hat R}&=\hat R\omega^\times-\sigma_R^\times\hat R,&
 \dot{\hat p}&=\hat v-\sigma_R^\times\hat p,\\
 \dot{\hat v}&=\hat g+\hat R a-\sigma_R^\times\hat v+K_vy,&
 \dot{\hat g}&=-\sigma_R^\times\hat g+K_gy.
\end{aligned}
\end{equation}
In the body coordinates $\hat v_\calB$, $\hat g_\calB$, and $\hat p_\calB:=\hat R^\top\hat p$, the terms in $\sigma_R$ cancel and the visual correction reduces to $Ly$, so the translational part of the observer is self-contained in body coordinates; this is the realization integrated in Section~\ref{sec:simresults}. 
This cancellation relies on the term $-\sigma_R^\times\hat p$ in the position dynamics, which results from the innovation $\Lambda\hat X$.
We now turn to the convergence of \eqref{eq:closederror}. Although the residual $y=C(t)x$ depends only on the velocity error, the coupling in $A(t)$ allows the gravity error to influence the output over time. 
Thus, convergence of both error components requires the visual constraints to supply sufficient information along the trajectory, in the following sense.

\begin{definition}\label{def:UO}
Let $A(t)$ and $C(t)$ be continuous and bounded on $\R_+$, and let $\Phi(s,t)$ be the transition matrix of $\dot z=A(t)z$, defined by $\partial_s\Phi(s,t)=A(s)\Phi(s,t)$ and $\Phi(t,t)=I$. The pair $(A(t),C(t))$ is uniformly observable if there exist $T,\alpha>0$ such that, for all $t\in\R_+$,
\begin{equation}\label{eq:gramian}
 \mathcal W(t,T):=\frac1T\int_t^{t+T}
 \Phi(s,t)^\top C(s)^\top C(s)\Phi(s,t)\,ds
 \succeq\alpha I.
\end{equation}
\end{definition}

For the present system, uniform observability can be characterized in terms of the motion and the scene geometry. Define the inertial bearings
\begin{equation}\label{eq:inertialbearing}
 q_i:=Rb_i=\frac{P_i-p}{\norm{P_i-p}},\qquad
 \dot q_i=Rs_i=-\frac{\pi_{q_i}v}{\norm{P_i-p}},
\end{equation}
and the matrix $N(t):=\sum_{i=1}^n(q_i\times\dot q_i)(q_i\times\dot q_i)^\top$. Each summand describes the direction constrained by one feature in inertial coordinates. The following result shows that a uniform lower bound on the time average of their sum is both necessary and sufficient for uniform observability of the six-dimensional error system.

\begin{proposition}\label{prop:trans}
Let Assumption~\ref{ass:standing} hold. The pair $(A(t),C(t))$ in \eqref{eq:Coutput} is uniformly observable if and only if there exist $\delta,\mu>0$ such that
\begin{equation}\label{eq:PE}
 \frac1\delta\int_t^{t+\delta}N(s)\,ds\succeq\mu I_3
 \qquad\text{for all }t\in\R_+.
\end{equation}
Under this condition, for any constant symmetric positive-definite weights $W_o,S$ and fixed initial matrix $P_0\succ0$, the origin of \eqref{eq:closederror} is globally uniformly exponentially stable.
\end{proposition}

\begin{proof}
Consider the homogeneous system $\dot z=A(t)z$ and express its two three-dimensional components in inertial coordinates by setting $\bar z:=T_R(t)z$, where $T_R:=\operatorname{diag}(R,R)$. Since $\dot R=R\omega^\times$, the transformed state and output matrices are $\bar A=\begin{bsmallmatrix}0&I_3\\0&0\end{bsmallmatrix}$ and $\bar C(t)=C(t)T_R(t)^\top$, and the identities $q_i\times\dot q_i=R\nu_i$ and \eqref{eq:rankoneflow} give $\bar C(t)^\top\bar C(t)=C_*^\top N(t)C_*$. Moreover, $N$ is continuous and bounded under Assumption~\ref{ass:standing}, since $\norm{q_i\times\dot q_i}\leq\norm{s_i}$. To relate the two observability conditions explicitly, let $\bar\Phi(s,t):=\exp(\bar A(s-t))$. The transition matrices satisfy $\Phi(s,t)=T_R(s)^\top\bar\Phi(s,t)T_R(t)$, and substituting this identity into \eqref{eq:gramian} yields $\mathcal W(t,T)=T_R(t)^\top\bar{\mathcal W}(t,T)T_R(t)$, where $\bar{\mathcal W}(t,T):=T^{-1}\int_t^{t+T}\bar\Phi(s,t)^\top C_*^\top N(s)C_*\bar\Phi(s,t)\,ds$. Because $T_R(t)^\top T_R(t)=I_6$, this change of coordinates preserves Euclidean norms. In particular, $\mathcal W(t,T)\succeq\alpha I_6$ if and only if $\bar{\mathcal W}(t,T)\succeq\alpha I_6$.

Suppose \eqref{eq:PE} holds. The pair $(\bar A,C_*)$ is observable, since $\operatorname{col}(C_*,C_*\bar A)=I_6$, and all eigenvalues of $\bar A$ are zero. Lemma~2.7 of \cite{hamel2017position}, applied with constant output matrix $C_*$ and bounded positive-semidefinite weight $N(t)$, therefore gives $T,\alpha>0$ such that $\bar{\mathcal W}(t,T)\succeq\alpha I_6$ for every $t\in\R_+$. This proves uniform observability of $(A(t),C(t))$. The horizon $T$ supplied by the lemma need not equal the excitation horizon $\delta$. Conversely, suppose $(A(t),C(t))$ is uniformly observable. The initial state $\bar z(t)=\operatorname{col}(\xi,0)$, with $\xi\in\R^3$, remains constant under $\dot{\bar z}=\bar A\bar z$, and evaluating the transformed Gramian on this state gives $T^{-1}\int_t^{t+T}\xi^\top N(s)\xi\,ds\geq\alpha\norm{\xi}^2$. Since this holds for every $\xi$, condition \eqref{eq:PE} follows with $\delta=T$ and $\mu=\alpha$.

For the Riccati equation, boundedness of the coefficients and $P_0\succ0$ ensure that its solution is defined and positive definite on $\R_+$ \cite[Lemma~2.5]{hamel2017position}. Let $\bar a:=\sup_t\norm{A(t)}$ and $\lambda_S:=\lambda_{\min}(S)$. The bound $\norm{\Phi(s,t)}\leq e^{\bar a(s-t)}$ for $s\geq t$ implies that the smallest singular value of $\Phi(t,s)=\Phi(s,t)^{-1}$ is at least $e^{-\bar a(s-t)}$. Thus the normalized process Gramian satisfies $T^{-1}\int_t^{t+T}\Phi(t,s)S\Phi(t,s)^\top ds\succeq\lambda_Se^{-2\bar aT}I_6$. The weighted observability Gramian also has a uniform positive lower bound, because $W_o\succeq\lambda_{\min}(W_o)I_{3n}$. Lemma~2.6 of \cite{hamel2017position} therefore yields constants $p_m,p_M>0$ such that
\begin{equation}\label{eq:Pbounds}
 p_m I_6\preceq P(t)\preceq p_M I_6
 \qquad\text{for all }t\in\R_+.
\end{equation}
These constants may depend on $P_0$, the coefficient and excitation bounds, and the chosen weights, but not on the initial state-estimation error. Finally, consider $V_x:=x^\top P^{-1}x$. Equations \eqref{eq:CRE} and \eqref{eq:closederror} give $\dot V_x=-x^\top(C^\top W_oC+P^{-1}SP^{-1})x\leq-(\lambda_S/p_M)V_x$. Together with $\norm{x}^2/p_M\leq V_x\leq\norm{x}^2/p_m$, this proves that for all $t\geq t_0\geq0$
\begin{equation}\label{eq:xrate}
 \norm{x(t)}\leq\sqrt{\frac{p_M}{p_m}}\,
 e^{-\lambda_S(t-t_0)/(2p_M)}\norm{x(t_0)},
\end{equation}
and establishes global uniform exponential stability.
\end{proof}

Condition \eqref{eq:PE} concerns the accumulation of information along the trajectory and separates the roles of the motion and the scene. By \eqref{eq:inertialbearing}, $q_i\times\dot q_i=(v\times q_i)/\norm{P_i-p}$, so that
\begin{equation}\label{eq:Nfactor}
 N=v^\times G\,(v^\times)^\top,\qquad
 G:=\sum_{i=1}^n\frac{q_iq_i^\top}{\norm{P_i-p}^2}.
\end{equation}
In particular, $N(t)v(t)=0$, meaning that the visual constraints provide no instantaneous information along the direction of motion. The speed can therefore only be recovered through the inertial dynamics and the time variation of the constraints. Pure rotation gives $N\equiv0$, while translation along a fixed inertial direction leaves that direction in the kernel of every $N(t)$, regardless of the camera rotation. Consequently, neither motion satisfies \eqref{eq:PE}.

The factorization also yields a necessary condition on the motion. If the feature ranges satisfy $\norm{P_i-p(t)}\geq\rho_*>0$ for all $i,t$, then $G(t)\preceq\bar\gamma I_3$ with $\bar\gamma:=n/\rho_*^2$, and since $v^\times(v^\times)^\top=\norm{v}^2I_3-vv^\top$, \eqref{eq:Nfactor} gives $N\preceq\bar\gamma(\norm{v}^2I_3-vv^\top)$. Averaging over a window and using \eqref{eq:PE}, the speed-weighted motion matrix must satisfy
\begin{equation}\label{eq:motionPE}
 \frac1\delta\int_t^{t+\delta}
 \big(\norm{v(s)}^2I_3-v(s)v(s)^\top\big)\,ds
 \succeq\frac{\mu}{\bar\gamma}I_3,
\end{equation}
so the inertial translation direction must vary over every window. If the speed is bounded above and away from zero, \eqref{eq:motionPE} is equivalent, up to constants, to the persistent excitation of the velocity direction required in \cite{bouazza2025observer}.

Conversely, a uniform condition on the scene geometry makes motion excitation sufficient. If $G(t)\succeq\gamma I_3$ for some $\gamma>0$, then $N\succeq\gamma(\norm{v}^2I_3-vv^\top)$, and persistent excitation of the speed-weighted motion matrix implies \eqref{eq:PE}. The scene condition requires at least three bearing directions spanning $\R^3$, with a lower bound that is uniform along the trajectory. It is sufficient but not necessary for \eqref{eq:PE}, since the time variation of the bearings can supply the missing directions, even with a single feature.

For example, consider $P_1=(0,0,h)^\top$ and the circular trajectory $p(t)=(\rho\cos\Omega t,\rho\sin\Omega t,0)^\top$, with $\rho>0$ and $\Omega\neq0$. Over one period $\delta=2\pi/|\Omega|$, direct calculation gives
\begin{equation}\label{eq:PEexample}
 \frac1\delta\int_t^{t+\delta}N(s)\,ds
 =\frac{\Omega^2}{(\rho^2+h^2)^2}
 \operatorname{diag}\!\left(\frac{\rho^2h^2}{2},\frac{\rho^2h^2}{2},\rho^4\right).
\end{equation}
Thus \eqref{eq:PE} holds if and only if the feature lies outside the plane of motion, $h\neq0$: one feature, despite contributing a single scalar constraint at each instant, can supply sufficient information through the variation of its bearing over time.

Condition \eqref{eq:PE} can be evaluated directly from the measurements without knowledge of the initial attitude, since a gyroscope-propagated rotation $R_g=QR$, with $Q\in\SO(3)$ constant, yields $N_g=QNQ^\top$, whose windowed averages have the same eigenvalues as those of $N$. The convergence analysis is established for the ideal coefficients; the effect of perturbed bearings, flow, and gyroscope on the homogeneous constraints is examined numerically in Section~\ref{sec:simresults}. Qualitatively, with perturbed coefficients $H_i^m=H_i+\Delta H_i$, the residual becomes $y_i=H_i^mx_v-\Delta H_iv_\calB$, and IMU perturbations similarly enter \eqref{eq:closederror} as additive inputs. As long as the perturbed pair remains uniformly observable, \eqref{eq:closederror} stays uniformly exponentially stable, so $x$ is ultimately bounded by a constant times the size of these inputs; the visual part scales with the speed $\norm{v_\calB}$.

\subsection{Reduced attitude and convergence of the cascade}

The convergent velocity--gravity subsystem provides the gravity estimate used by the attitude correction. Because rotations about gravity remain unobservable, the relevant attitude error is the discrepancy in the gravity direction. To analyze this error, define $r:=\tilde R e_3\in\mathbb S^2$ and $u:=Rx_g/g_0$. The variable $r$ measures the reduced-attitude error, since $\norm{r-e_3}=\norm{R^\top e_3-\hat R^\top e_3}$, so $r=e_3$ if and only if $\hat R^\top e_3=R^\top e_3$, while $u$ is the normalized gravity error in inertial coordinates. From \eqref{eq:xdef}, $J\hat Xe_g=\hat g=\tilde R^\top(g-Rx_g)$, and substitution into \eqref{eq:attitudeobserver} and \eqref{eq:groupcolumns} gives
\begin{equation}\label{eq:reducedexact}
 \dot r=k_g\pi_r(e_3-u).
\end{equation}
For fixed $x(0)$ and $P_0$, the input $u$ is determined by \eqref{eq:closederror} and the true trajectory and does not depend on $r(0)$, so \eqref{eq:closederror}, \eqref{eq:reducedexact} form a cascade. When $u\equiv0$, $V_r:=1-e_3^\top r$ satisfies $\dot V_r=-k_gV_r(2-V_r)$, so $e_3$ is asymptotically stable and $-e_3$ is the only other equilibrium. Since $u$ need not be identically zero, this autonomous argument does not apply directly; the next result accounts for the input.

\begin{theorem}\label{thm:cascade}
Let Assumption~\ref{ass:standing} and \eqref{eq:PE} hold, and let $k_g>0$ and $W_o,S,P_0\succ0$, with $P_0$ independent of the initial estimate. Then the observer \eqref{eq:groupobserver}--\eqref{eq:attitudeobserver}, with the gain \eqref{eq:gainlift} and Riccati equation \eqref{eq:CRE}, is forward complete, and its velocity--gravity error satisfies \eqref{eq:xrate}. The equilibrium $(x,r)=(0,e_3)$ of \eqref{eq:closederror}, \eqref{eq:reducedexact} is locally uniformly exponentially stable, and the initial conditions in $\R^6\times\mathbb S^2$ from which it is not approached form a set of zero measure. In particular, it is almost globally asymptotically stable.
\end{theorem}
\begin{proof}
We first verify that the observer solutions are defined for all positive times. By Proposition~\ref{prop:trans} and its proof, $P$ satisfies \eqref{eq:Pbounds} and $x$ satisfies \eqref{eq:xrate}, so $u$ is bounded and decays exponentially. The variables $r$ and $\hat R$ evolve on compact sets, $\hat v=\hat R(v_\calB-x_v)$ and $\hat g=\hat R(g_\calB-x_g)$ are bounded on compact time intervals, and the position equation in \eqref{eq:observercomponents} is linear in $\hat p$; hence the observer is forward complete.

To establish local exponential stability, let $\chi(r):=1-(e_3^\top r)^2=V_r(2-V_r)$. Along \eqref{eq:reducedexact},
\begin{equation}\label{eq:reduceddirectV}
 \dot V_r=-k_g \chi(r)+k_g e_3^\top\pi_r u.
\end{equation}
Since $\norm{\pi_r e_3}^2=\chi(r)$ and $\chi(r)\geq V_r$ when $V_r\leq1$, Young's inequality gives, on $V_r\leq1$,
\begin{equation}\label{eq:localattitude}
 \dot V_r\leq-\frac{k_g}{2}V_r+\frac{k_g}{2g_0^2}\norm{x}^2.
\end{equation}
Let $c_x:=k_g p_M^2/(g_0^2\lambda_S)$ and $V:=V_r+c_xV_x$. Since $x^\top P^{-1}SP^{-1}x$ is bounded below both by $\lambda_S\norm{x}^2/p_M^2$ and by $\lambda_SV_x/p_M$, half of the dissipation in $c_x\dot V_x$ cancels the last term of \eqref{eq:localattitude}, and the other half gives
\begin{equation}\label{eq:cascadeLyapunov}
 \dot V\leq-\min\!\left\{\frac{k_g}{2},\frac{\lambda_S}{2p_M}\right\}V
 \qquad\text{when }V_r\leq1.
\end{equation}
The sublevel set $\{V<1\}$ is therefore forward invariant. Since $2V_r=\norm{r-e_3}^2$ and \eqref{eq:Pbounds} holds uniformly, \eqref{eq:cascadeLyapunov} proves local uniform exponential stability. We next consider arbitrary initial conditions. The last term in \eqref{eq:reduceddirectV} is integrable on $\R_+$ because $u$ decays exponentially. Integrating \eqref{eq:reduceddirectV} shows that $\chi(r(\cdot))$ is integrable and that $V_r(t)$ has a limit. Since $\dot r$ is bounded, $\chi(r(\cdot))$ is uniformly continuous, and Barbalat's lemma gives $\chi(r(t))\to0$. Because $V_r$ converges, every solution converges to $e_3$ or to $-e_3$. It remains to show that convergence to $-e_3$ is exceptional. Fix $x(0)$, and hence $u(\cdot)$. Along two solutions $r_1,r_2$ of \eqref{eq:reducedexact} driven by this input, $d:=\norm{r_1-r_2}^2$ satisfies $\dot d=c(t)\,d$ with $c:=-k_g(e_3-u)^\top(r_1+r_2)$. If both converged to $-e_3$, then $c(t)\to2k_g>0$, so $d$ would grow exponentially; but $d\leq4$ on the unit sphere, so $d(0)=0$. Hence, for each $x(0)$, at most one initial reduced attitude leads to $-e_3$. The set of initial conditions $(x(0),r(0))$ converging to $-e_3$ is measurable, being the set where $e_3^\top r(k)<0$ for all large integers $k$, a countable combination of open sets by continuous dependence on initial conditions; since it meets each $\{x(0)\}\times\mathbb S^2$ in at most one point, it has zero measure by Fubini's theorem.
\end{proof}

Theorem~\ref{thm:cascade} concerns the gauge-invariant quantities of Section~\ref{sec:problem}. For the other blocks of $E$, for almost every initial condition $(x,r)$ eventually enters the sublevel set $\{V<1\}$ used in the proof, after which $r-e_3$ and $u$ decay exponentially. Since $\sigma_R=k_g\,e_3\times\tilde R^\top(e_3-u)$ and $\norm{e_3\times\tilde R^\top e_3}=\norm{r\times e_3}\leq\norm{r-e_3}$, it follows that $\sigma_R$ is integrable and $\tilde R$ converges to some $\tilde R_\infty$ satisfying $\tilde R_\infty e_3=e_3$. The attitude is therefore recovered up to a constant yaw offset. Furthermore, since $\dot{\tilde p}=Rx_v$ is integrable, $\tilde p$ converges to some $\tilde p_\infty$, so $E(t)\to\mathcal G_{\tilde R_\infty,\tilde p_\infty}\in\mathcal S$. Thus, the right-invariant error converges to an element of the gauge group, while the raw position error $p-\hat p=(I_3-\tilde R^\top)p+\tilde R^\top\tilde p$ need not vanish or remain bounded.

\section{Simulation results}\label{sec:simresults}

The numerical study uses prescribed rigid-body kinematics to generate spherical bearings and their exact derivatives. All features remain visible throughout a run, with no field-of-view limits or occlusions, and no vehicle-dynamics model or image-level optical-flow estimator is involved. The simulations separate ideal measurements under motions with and without excitation, recovery from randomized initial errors, and odometry from a correct initialization under perturbed measurements. 
All runs integrate \eqref{eq:observercomponents} over $40\,\mathrm{s}$, with the translational part implemented in the body coordinates $\hat v_\calB$ and $\hat g_\calB$, with parameters $P_0=I_6$, $S=30I_6$, $W_o=5I_{3n}$, and $k_g=1.5\,\mathrm{s}^{-1}$. Each feature keeps the same weight as $n$ varies. 
 Further runs are recorded in an online viewer, \url{https://sberkane.github.io/vio-lab/}.
 
\subsection{Ideal measurements and excitation}
The circle, figure-eight, and straight trajectories are, in metres with $t$ in seconds, $p_c=(3\cos0.5t,3\sin0.5t,0)^\top$, $p_f=(3\cos0.5t,2\sin t,1.2\sin0.35t)^\top$, and $p_s=(-6+0.3t,-1,0.5)^\top$. The attitude is $R=R_z(\psi)R_y(\theta)R_x(\phi)$, where $R_j$ denotes rotation about axis $j$ and 
angles are in radians. For the figure-eight, $\phi=0.16\sin(0.6t)$, $\theta=0.13\cos(0.45t)$, and $\psi=0.28\sin(0.25t)+0.15t$; for the circle, $\phi=0.12\sin(0.6t)$, $\theta=0.10\cos(0.4t)$, and $\psi=0.5t+0.3$; for straight translation, $\phi=0.12\sin(0.7t)$, $\theta=0.10\cos(0.35t)$, and $\psi=0.25\sin(0.45t)+0.5$.
The IMU signals are $a=R^\top(\ddot p-g)$ and $\omega^\times=R^\top\dot R$, and bearings and flows follow \eqref{eq:bearing}--\eqref{eq:flow}. The scene contains $P_1=(0,0,4)^\top\,\mathrm m$ and nineteen points with horizontal radii in $[5,8]\,\mathrm m$ and third coordinates in $[2,6]\,\mathrm m$; the observer receives no ranges or landmark positions.
The initial errors are $R(0)x_v(0)=(-1.5,1,-0.6)^\top\,\mathrm{m/s}$, $R(0)x_g(0)=(0.35,-0.2,0.5)^\top\,\mathrm{m/s^2}$, $\hat R_0=\exp(\zeta_0^\times)R_0$ with $\zeta_0=(40,-30,55)^\top\pi/180$, and $\hat p_0-p_0=(0.8,-1,0.5)^\top\,\mathrm m$. Ideal runs use DOP853 integration with relative and absolute tolerances $2\times10^{-9}$ and $2\times10^{-11}$ and a maximum step of $0.05\,\mathrm s$.
 We report $\norm{x_v}$, $\norm{x_g}$, and the reduced-attitude angle $\theta_g:=\arccos(e_3^\top\tilde R e_3)$ in degrees.

\begin{figure}[t]
 \centering
 \includegraphics[trim={0cm 0cm 0cm 0cm}, width=.6\columnwidth]{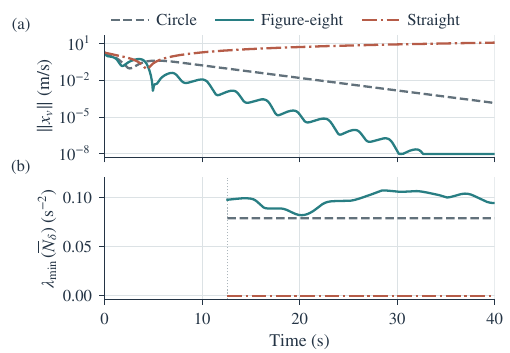}
 \caption{Ideal measurements, $n=8$: velocity-error norm and minimum eigenvalue of $\overline N_\delta(t):=\delta^{-1}\int_{t-\delta}^{t}N(s)\,ds$, with $\delta=4\pi\,\mathrm s$ (complete windows only). Velocity errors are floored at $10^{-8}\,\mathrm{m/s}$ for display. Straight translation violates the excitation condition despite the changing attitude.}
 \label{fig:sim-ideal}
\end{figure}

Figure~\ref{fig:sim-ideal} compares the motions for $n=8$. The excitation matrix is positive definite for both the circle and figure-eight, but singular under straight translation. The along-track inertial errors satisfy $(Rx_v)_1=-1.5+0.35t$ and $(Rx_g)_1=0.35$ independently of $n$, so adding features cannot remove this mode. For the circle, one feature reproduces \eqref{eq:PEexample}, with minimum eigenvalue $0.0288\,\mathrm{s}^{-2}$ over a $4\pi\,\mathrm s$ window. Convergence is nevertheless slow, with $\norm{x_v(40)}=0.0761\,\mathrm{m/s}$ for one feature and $1.44\times10^{-4}\,\mathrm{m/s}$ for eight. For the figure-eight, $q_1\times\dot q_1$ at $t=0,10\pi,20\pi$ span $\R^3$ and the translation has period $40\pi\,\mathrm s$, so \eqref{eq:PE} holds with $\delta=40\pi\,\mathrm s$; the $4\pi\,\mathrm s$ windows of Fig.~\ref{fig:sim-ideal} provide a shorter diagnostic.

\begin{figure}[t]
 \centering
 \includegraphics[trim={0cm 0cm 0cm 0cm}, width=.6\columnwidth]{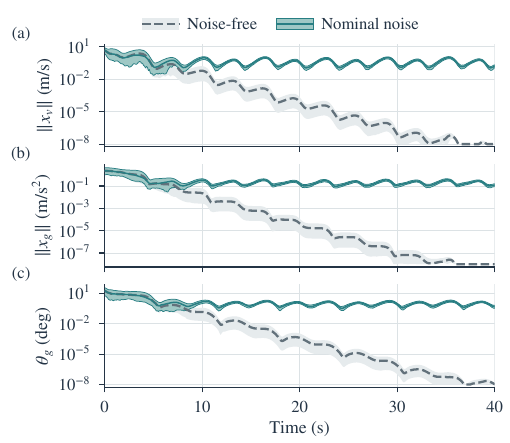}
 \caption{Recovery on the figure-eight, $n=8$: 100 randomized initial states per condition, noise-free and at nominal noise. Curves are pointwise medians, bands empirical 5th--95th percentiles; values below $10^{-8}$ are floored. Reduced attitude excludes the unobservable yaw.}
 \label{fig:sim-recovery}
\end{figure}

\begin{figure}[t]
 \centering
 \includegraphics[trim={0cm 0cm 0cm 0cm}, width=.6\columnwidth]{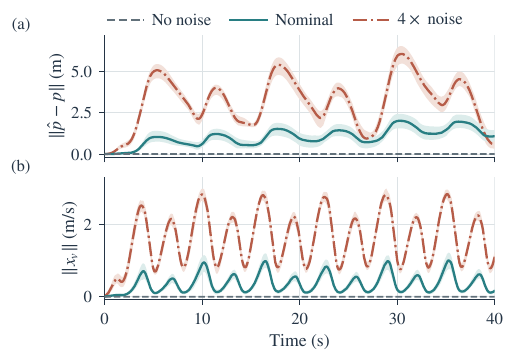}
\caption{Matched-initialization odometry on the figure-eight, $n=8$, at nominal and fourfold noise (100 trials each) and without noise. Position error is in the common initial frame, unaligned.}
 \label{fig:sim-noise}
\end{figure}

\subsection{Measurement perturbations and recovery}
 
Each perturbation component is the random signal
$\epsilon(t)=(\sigma/\sqrt8)\sum_{k=1}^{8}\big(a_k\cos(2\pi f_kt)+b_k\sin(2\pi f_kt)\big)$, with standard Gaussian coefficients drawn independently across harmonics, components, features and sensors, so that $\epsilon(t)$ has standard deviation $\sigma$ at each $t$. The frequencies are geometrically spaced over $0.2$--$2\,\mathrm{Hz}$ for the bearings and $0.2$--$5\,\mathrm{Hz}$ for the IMU. The nominal values of $\sigma$ are $0.001$ for the bearings, $0.002\,\mathrm{rad/s}$ for the gyroscope, and $0.02\,\mathrm{m/s^2}$ for the accelerometer. The second noise level uses perturbations with four times the nominal amplitudes. With $z_i=b_i+\eta_i$, the camera signals are $b_i^m=z_i/\norm{z_i}$ and $\dot b_i^m=\pi_{b_i^m}(\dot b_i+\dot\eta_i)/\norm{z_i}$, so the measured bearing and its derivative are consistent, and the rotational compensation uses the perturbed gyroscope. The resulting $H_i^m$ need not satisfy $H_i^m v_{\cal B}=0$. IMU biases are not included.

Two initializations distinguish recovery from odometry drift. The recovery trials draw independent $\delta v\sim\mathcal N(0,3^2I_3)$ in $\mathrm{m/s}$, $\delta g\sim\mathcal N(0,1.5^2I_3)$ in $\mathrm{m/s^2}$, and $\delta\theta\sim\mathcal N(0,(15\pi/180)^2I_3)$ in radians, and set $\hat R_0=\exp(\delta\theta^\times)R_0$, $\hat v_{{\cal B},0}=R_0^\top(v_0+\delta v)$, and $\hat g_{{\cal B},0}=R_0^\top(g+\delta g)$. For odometry, all initial estimates equal the truth. Both studies shift the trajectory and landmarks so that $p_0=\hat p_0=0$, retain the position equation of \eqref{eq:observercomponents}, and never align the estimated path. Each noisy condition uses 100 paired trials, with a deterministic noise-free reference.
Fourth-order Runge--Kutta integration is used with a $0.01\,\mathrm s$ step.
 Figures~\ref{fig:sim-recovery}--\ref{fig:sim-noise} show the figure-eight with eight features.
 Without noise, the median velocity-error norm at $40\,\mathrm s$ is $1.20\times10^{-8}\,\mathrm{m/s}$.
 With measurement perturbations, errors persist after the transient. Position time-RMS is computed separately for each trial as $\big(40^{-1}\int_0^{40}\norm{p-\hat p}^2dt\big)^{1/2}$. With matched initialization, its median is $1.132\,\mathrm m$ at nominal noise and $3.489\,\mathrm m$ at elevated noise. The final velocity-error medians are $0.158$ and $1.107\,\mathrm{m/s}$.
 The matched-initialization runs also reveal a reduction in estimated speed under measurement perturbations. Over $t\geq5\,\mathrm s$, the median across trials of the time-averaged ratio $\norm{\hat v_\calB}/\norm{v_\calB}$ is $0.78$ at nominal noise and $0.31$ at elevated noise, whereas the RMS error component orthogonal to $v_\calB$ has median $0.045$ and $0.18\,\mathrm{m/s}$.
 In these runs with fixed per-feature weights, the estimated speed decreases as $n$ increases: at nominal noise and $n=4,8,20$, the speed-ratio medians are $0.87,0.78,0.56$, and the position time-RMS medians are $0.645,1.132,1.871\,\mathrm m$. This bias arises because $H_i^mv_\calB\neq0$, the visual innovation, which drives each $H_i^m\hat v_\calB$ to zero, pulls $\hat v_\calB$ toward zero, increasingly so as the total visual weight grows with $n$. Thus, greater ideal excitation does not guarantee better accuracy under perturbations.
 These finite-horizon tests illustrate measurement sensitivity. The convergence claims remain those of the ideal model.
 
\section{Conclusions}\label{sec:conclusion}

Individual bearing-flow constraints can serve directly as implicit outputs of a Riccati observer on the gravity-augmented group $\SEthree$, without landmark estimation or reconstruction of the direction of motion. The auxiliary gravity state decouples the body-frame velocity--gravity error from the attitude correction, and its uniform observability is equivalent to persistent excitation of the inertial bearing-flow matrix. Under this condition the velocity--gravity error converges uniformly exponentially, the reduced attitude is almost globally recovered, and position and yaw remain gauge freedoms. Simulations confirm the role of excitation but show lower speed estimates under measurement perturbations, increasingly so with the number of equally weighted features. Constraint weightings that mitigate this effect, IMU bias estimation, and changing feature sets remain open.

\section*{Acknowledgment}
This work was supported by the ``Grands Fonds Marins'' Project Deep-C, the ASTRID ANR project ASCAR, and also in part by NSERC-DG RGPIN-2020-04759 and Fonds de recherche du Qu\'ebec (FRQ).

The authors used ChatGPT (OpenAI) and Claude (Anthropic) for copy-editing and assistance with the simulation code in Section \ref{sec:simresults}.
All AI-assisted content was reviewed and verified by the authors, who take full responsibility for this paper.

\bibliographystyle{unsrtnat}
\bibliography{ref}

\end{document}